\documentclass[useAMS,usenatbib]{aa}
\usepackage[english]{babel}
\usepackage{amsmath,amssymb}

\usepackage{graphicx}

\usepackage[normalem]{ulem}

\usepackage{verbatim}
\usepackage{color}
\usepackage{multirow}
\usepackage{mathtools}
\usepackage{epstopdf}

\usepackage{hyperref}
\usepackage{url}
\usepackage{xcolor}

\usepackage{gensymb}

\def\nh2{n_{\rm H2}}
\def\fh2{f_{\rm H2}}

\def\angstrom{\textrm{A\kern -1.3ex\raisebox{0.6ex}{$^\circ$}}}

\bibpunct{(}{)}{;}{a}{}{,}             
\definecolor{cobalt}{rgb}{0.06, 0.2, 0.65}
\hypersetup{
  colorlinks,
  citecolor=cobalt,
  linkcolor=[rgb]{0.8, 0.2, 1.0},
  urlcolor=cobalt,
}
\usepackage{appendix}
\usepackage{verbatim}
\usepackage{enumitem}

\begin{document}

   \title{A Selection Aware View of Black Hole–Galaxy Coevolution at High Redshift}

   \author{F. Ziparo
          \inst{1}
          \and
          S. Carniani
          \inst{1}
          \and
          S. Gallerani
          \inst{1}
          \and
          B. Trefoloni
          \inst{1}
          }

   \institute{Scuola Normale Superiore,
              Piazza dei Cavalieri 7, I-56126 Pisa, Italy\\
              \email{francesco.ziparo@sns.it}
        }

   \date{}

 
  \abstract{

The large population of broad-line Active Galactic Nuclei (AGN) observed with the James Webb Space Telescope (JWST) at $z \gtrsim 4$ opens a new window onto the black hole–galaxy connection in the first Gyr of cosmic history. We use the JADES survey-level dataset and develop a forward-modeling Bayesian framework that explicitly accounts for broad H$\alpha$ detectability, ensuring that selection effects are incorporated into the likelihood function. With this approach, we constrain the black hole–stellar mass ($M_{\mathrm{BH}}$–$M_\star$) relation to be $\log M_{\rm BH} = -4.06^{+0.50}_{-0.51} + 1.17^{+0.06}_{-0.06}\,\log M_\star$, with an intrinsic orthogonal scatter of $\sigma_{\rm int} = 0.63^{+0.14}_{-0.11}$ dex. The slope and normalization are consistent with local determinations, indicating that the average scaling was already established by $z \sim 4$–6. This suggests that the primary evolution of the relation occurs in its dispersion rather than in its mean normalization. In contrast, the substantially larger intrinsic scatter relative to the nearby Universe reveals a wider diversity of black hole–galaxy growth histories, likely driven by bursty accretion, delayed feedback, and differences in merger or seeding histories. Future JWST samples will be crucial to test whether this increased scatter is a persistent feature of the high-redshift Universe.

}
   \maketitle
%

\section{Introduction}

Supermassive black holes (SMBHs), with masses ranging from $10^6$ to over $10^{10}\,M_\odot$, are ubiquitously found at the centers of massive galaxies \citep{KH13, RV15} both in the local and high-$z$ Universe. A wealth of observational evidence has established that the growth of SMBHs is closely linked to the formation and evolution of their host galaxies. This co-evolution is supported by tight empirical correlations between SMBH mass and large-scale galactic properties such as stellar mass, bulge luminosity, and velocity dispersion \citep{Magorrian98, Ferrarese00, Gebhardt00}. These correlations are thought to originate from a combination of interconnected physical mechanisms that operate during the active phases of galaxy and black hole growth. In particular, both the SMBH and its host galaxy are believed to accrete gas from a common reservoir, resulting in a coupled evolution of their mass assembly histories \citep{DiMatteo05, Hopkins06}. Simultaneously, feedback from the active galactic nucleus (AGN) can inject energy and momentum into the surrounding interstellar medium, in the form of radiation, winds, or relativistic jets, thereby regulating star formation and SMBH accretion \citep{Harrison17}. This self-regulating cycle is expected to play a key role in shaping the observed scaling relations and to drive the joint evolution of SMBHs and galaxies over cosmic time.

Although the scaling relations between SMBHs and their host galaxies are well established in the local Universe, their evolution at earlier cosmic epochs remains poorly constrained. Different studies have begun to trace these relations to intermediate redshifts ($z\sim1-3$), finding hints of mild evolution, with SMBHs appearing somewhat overmassive relative to their hosts at fixed stellar mass \citep{Merloni10,Bennert11,Ding20}. However, observational limitations and selection biases still hamper a consistent picture across cosmic time, especially beyond the peak of BH and galaxy growth. 
In particular, it is still unclear whether the same correlations hold at $z \gtrsim 4$, when both BH accretion and star formation were rapidly evolving \citep{Madau14,Aird15}. With a few noticeable exceptions (e.g. \citealt{higzDynamBHmass}), at these redshifts direct dynamical measurements of SMBH masses are not feasible, as the angular resolution required to resolve stellar or gas kinematics in galaxy nuclei exceeds current instrumental capabilities \citep[e.g.,][]{KH13,Greene20}. 
Instead, BH masses are typically inferred using single-epoch virial estimators based on the widths of broad emission lines and the continuum luminosity, under the assumption that the gas motions in the broad-line region (BLR) are dominated by the gravitational potential of the SMBH \citep{Vestergaard06,Shen13}. 
These virial relations are calibrated against local reverberation-mapped AGN \citep{Kaspi00, Cho23, Dallabonta2025}, yet their validity at high redshift remains uncertain due to possible changes in BLR structure, ionization conditions, and accretion properties. Among the available diagnostics, the H$\alpha$ line is of particular interest because it provides one of the most reliable virial mass estimators, being less affected by non-virial components and dust extinction compared to UV lines such as \ion{C}{IV} or \ion{Mg}{II} \citep[e.g.,][]{GreeneHo05}. 
However, detecting and accurately decomposing the broad H$\alpha$ emission at $z>4$ is challenging, since the limited spectral quality often hampers a clean separation between the broad and narrow components \citep{Carnall23, Matthee24}. As a result, current samples may be biased toward the most luminous $L_{\mathrm{H}\alpha}\gtrsim 10^{42}\ \rm erg\ s^{-1}$, high-accretion SMBHs $\lambda_{\rm Edd}\gtrsim 0.5$ \citep{Maiolino24}, while a large fraction of the AGN population could remain below the detection threshold.  Flux-limited selection further amplifies these effects through Lauer bias \citep{Lauer07}, potentially distorting the observed $M_{\mathrm{BH}}$-$M_\star$ distribution relative to the intrinsic one. Quantifying and correcting for these selection effects is therefore essential to recover unbiased scaling relations and to trace SMBH-galaxy coevolution at cosmic dawn.

JWST observations now enable rest-frame optical spectroscopy of AGN at $z>4$ \citep{JWST}, allowing simultaneous measurements of host stellar masses and virial BH masses from broad H$\alpha$ and H$\beta$ emission. Several studies have therefore begun to place the first empirical constraints on the $M_{\mathrm{BH}}$–$M_\star$ relation at early cosmic times using spectroscopically confirmed broad-line AGN samples \citep{Harikane23, Kocevski23, Maiolino24, Matthee24, Judobaliz25}. However, these samples remain heterogeneous and are often limited by small-number statistics, spectral quality, and observational biases that favor the most luminous sources. 

A recent attempt to address this challenge was presented by \citet{Pacucci23}, who compiled a sample of spectroscopically confirmed broad-line AGN at $z \gtrsim 4$ and derived a $M_{\mathrm{BH}}$–$M_\star$ relation while accounting for basic observability thresholds. Their analysis suggests that SMBHs at high redshift appear overmassive compared to local expectations, potentially hinting at early rapid BH growth. However, the limited sample size (20 objects) may restrict the statistical robustness of the inferred relation, and the treatment of selection effects—while innovative—may not fully capture the complex, multivariate detectability of broad-line components. Building on this approach, a similar truncated-likelihood analysis applied to a substantially larger JWST-selected sample \citep{Jones+25} yielded results broadly consistent with that study, suggesting that the inferred offset is not driven solely by limited statistics but may depend sensitively on the adopted treatment of selection effects. Other studies have proposed that the seeming overmassiveness may instead result from observational biases alone \citep{Li25}. Similarly, population-based modeling of high-redshift quasar samples has shown that a normalization consistent with the local relation, combined with a large intrinsic scatter, can naturally reproduce the observed AGN population without invoking systematically overmassive black holes \citep{Silverman+25}. Complementary stacking analyses \citep{Geris25} have revealed previously undetected low-mass black holes with typical stellar-to-black hole mass ratios of $\sim10^{-3}$, consistent with local scaling relations \citep{RV15, Greene20}. These detections indicate that current flux-limited samples miss part of the population and further motivate a careful treatment of selection biases. Such a framework is essential to recover unbiased estimates of the scaling relation from flux-limited high-redshift samples.

In this work, we develop a forward-modeling framework to infer the $M_{\mathrm{BH}}$–$M_\star$ relation at $z \gtrsim 4$ while accounting for selection effects in broad-line AGN samples. In Sec. \ref{sec:methods}, we describe the construction of mock H$\alpha$ spectra based on physical input parameters, and the derivation of detectability maps through spectral fitting under realistic noise conditions. These maps serve as selection functions in a truncated-likelihood formalism, which we use to fit a parametric $M_{\mathrm{BH}}$–$M_\star$ relation to a JWST-selected sample of broad-line AGN. Our results are presented in Sec. \ref{sec:results}, where we assess whether the observed offset from the local relation reflects a true evolution or is consistent with observational biases. Finally, in Sec. \ref{sec:discussion}, we discuss the implications of our findings in the broader context of SMBH–galaxy coevolution, including recent JWST results, current models of early black hole growth, and the observational strategies needed to further constrain the scaling relations at high redshift.

\section{Methods}
\label{sec:methods}
In this section, we describe the procedure adopted to generate synthetic H$\alpha$ spectra and to assess the detectability of AGN signatures under varying physical conditions. Throughout this analysis, we consider only unobscured, broad-line (type~1) AGN, excluding sources where the broad component is hidden by nuclear obscuration. Our approach is based on a forward-modeling framework in which the emission-line profiles are derived analytically from the underlying galaxy and BH properties, allowing us to quantify how observational selection affects the recovery of broad-line components. In practice, we define the bias as the variation in detection probability of the broad H$\alpha$ component across the ($\log M_\star$, $\log M_{\mathrm{BH}}$) plane, which is the parameter space where the intrinsic scaling relation is measured. The bias is computed by sampling a four-dimensional grid in black hole mass, stellar mass, Eddington ratio, and star formation rate (see Sec.~\ref{sec:parameterspace} for ranges), and evaluating for each point the fraction of realizations in which a broad component is recovered under realistic noise conditions. This provides an empirical estimate of the selection function, which we later incorporate into a truncated-likelihood model to correct the inferred $M_{\mathrm{BH}}$–$M_\star$ relation for incompleteness. Although this estimate is necessarily approximate, it captures the dominant dependence of detectability on host and BH properties, and its validity is verified through recovery tests presented in Sec.~\ref{sec:recovery_test}. By explicitly modeling this bias, we can assess whether the apparent lack of low-mass BHs at high redshift reflects a true physical evolution or results from observational selection effects.

\subsection{Mock H$\alpha$ spectra}
\label{sec:mock}

\begin{figure}
\centering
\includegraphics[width=0.5\textwidth]{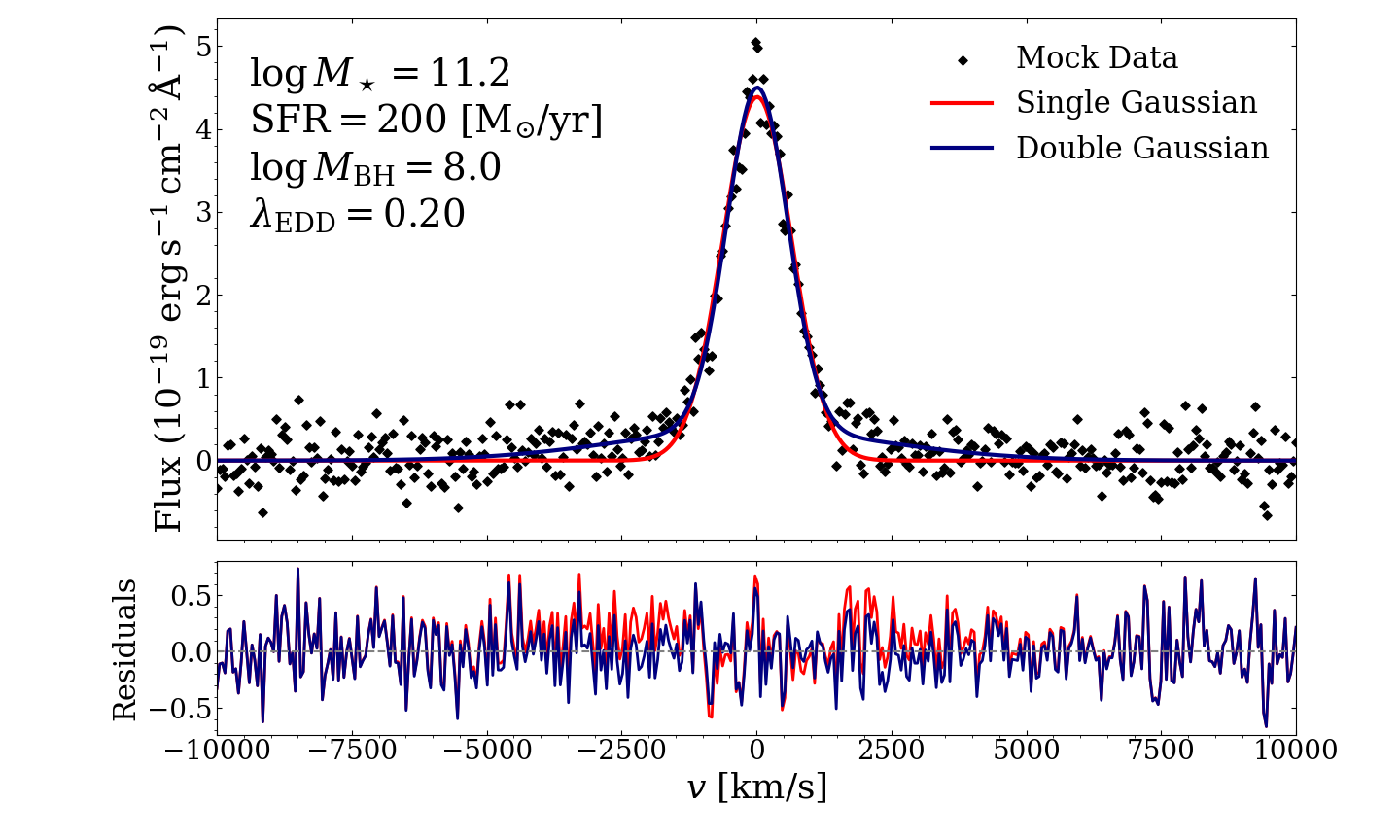}
\caption{
Example of spectral fitting for a mock H$\alpha$ emission line. 
\textbf{Top panel:} synthetic data (black diamonds) overlaid with the single-Gaussian fit (red line) and the double-Gaussian fit (blue line). 
\textbf{Bottom panel:} residuals for the single-Gaussian (red) and double-Gaussian (blue) models, normalized to the noise RMS. 
In this case, the broad component is correctly identified with $\Delta{\rm BIC} = 52$. 
}
\label{fig:fit_example}
\end{figure}

We generated mock emission-line spectra centered on the H$\alpha$ line (6563 \textrm{\AA}) following a procedure similar to the one outlined in \citet[][see their Sec. 5]{Srija25}. We model both the narrow and the broad component as Gaussian functions. The physical properties of the two components were computed directly from four input free parameters:  black hole mass ($M_{\mathrm{BH}}$),  black hole accretion rate in units of the Eddington rate ($\lambda_{\mathrm{Edd}}$),  stellar mass of the host galaxy ($M_\star$), and  star formation rate (SFR).

To determine the narrow component, we assumed that its luminosity scales linearly with the star formation rate of the host galaxy following the calibration by \citep{Kennicutt12}:

\begin{equation}
L^{\mathrm{narrow}}_{\mathrm{H}\alpha} = 1.86 \times 10^{41} \times \mathrm{SFR} \quad \mathrm{erg\,s^{-1}}.
\end{equation}

We derived the width of the narrow component from the gas velocity dispersion, $\sigma_{\rm dyn}$, computed from the dynamical mass of the host galaxy. Following previous studies \citep{Maiolino24,Uebler24}, we adopted:
\begin{equation}
M_{\mathrm{dyn}} = \frac{K(n)\,K(q)\,\sigma_{\mathrm{dyn}}^2\,R_{\mathrm{eff}}}{G},
\end{equation}
which we inverted to obtain $\sigma_{\rm dyn}$. The effective radius $R_{\mathrm{eff}}$ was derived from the stellar mass using the empirical $M_\star$--$R_{\mathrm{eff}}$ relation by \citet{Allen25}. We fixed the structural parameters to $K(n) = 8.06$ and $K(q) = 1.02$, corresponding to a Sérsic index of $n = 1$ and an axis ratio of $q \approx 0.73$, consistent with typical star-forming disk galaxies at high redshift \citep[e.g.,][]{vanderWel22}. The resulting dispersion $\sigma_{\mathrm{dyn}}$ was then used as the Gaussian standard deviation of the narrow mock H$\alpha$ component in each mock spectrum.

To model the broad component, we adopted the empirical relation used to study the emission of the broad-line regions in AGN. The integrated flux of the line was based on the line luminosity  derived from the optical continuum luminosity linking the 5100\,\AA\ monochromatic luminosity and the broad H$\alpha$ luminosity, as defined in \citep{Reines13}:
\begin{equation}
L_{\rm H\alpha}^{\rm broad}=5\times 10^{42}\biggl(\frac{L_{5100}}{10^{44}\rm erg\ s^{-1}}\biggl)^{1.157}\ \rm erg\ s^{-1},
\end{equation}
where the 5100\,\AA\ luminosity was computed as $L_{\mathrm{5100}} = f_{\mathrm{bol}} \cdot L_{bol}$, assuming a bolometric correction factor $f_{\mathrm{bol}} = 0.1$ \citep{Lusso12}. The bolometric luminosity was estimated from the BH mass and Eddington ratio as 
$L_{\mathrm{bol}} = \lambda_{\mathrm{Edd}} \, L_{\mathrm{Edd}} = \lambda_{\mathrm{Edd}} \, (1.26 \times 10^{38} \, M_{\mathrm{BH}} / M_\odot)\,\mathrm{erg\,s^{-1}}$.
 The width of the broad component (FWHM) was obtained by inverting the virial relation \citep{Reines13}:

\begin{align}
\log \left( \frac{M_{\mathrm{BH}}}{M_\odot} \right) &= 6.6 
+ 0.47 \log \left( \frac{L^{\mathrm{broad}}_{\mathrm{H}\alpha}}{10^{42}\,\mathrm{erg\,s^{-1}}} \right) \nonumber \\
&\quad + 2.06 \log \left( \frac{\mathrm{FWHM}_{\mathrm{broad}}}{10^3\,\mathrm{km\,s^{-1}}} \right).
\end{align}
To simulate NIRSpec data observations with a spectral resolution  of $\Delta\lambda/\lambda\approx 2700$, each synthetic spectrum was convolved with the line spread function of the instrument  and sampled with a spectral channel of  40 $\rm km/s$.   We then added Gaussian white noise to each synthetic spectrum. We generated two datasets:  a deep dataset with a noise level of $6 \times 10^{-21}$ erg\,s$^{-1}$\,cm$^{-2}$\,\AA$^{-1}$, matching the depth of  the deepest observations in the JADES survey \citep{Maiolino24, CurtisLake25, Scholtz25}, and a shallower dataset with a noise level of $2.4 \times 10^{-20}$ erg\,s$^{-1}$\,cm$^{-2}$\,\AA$^{-1}$, representative of typical JWST observations to date, used to assess the effect of increased noise on the detectability of faint broad emission-line components. The latter case was included to explore a more conservative observational scenario, since the JADES survey generally provides high signal-to-noise spectra, while our compiled sample spans a broader range of data quality and is intended to support a methodology applicable to heterogeneous JWST datasets.

\subsection{Spectral fitting and broad–line detectability}
\label{sec:fitting}

Following the same approach adopted in observational studies, each mock  $H\alpha$ spectrum was fitted with two line profile models:  
(i) a \textit{single–Gaussian} profile that represents the narrow component alone, and  
(ii) a \textit{double–Gaussian} profile that adds a putative broad component.  

The fits were performed with the \texttt{scipy.optimize.curve\_fit} routine, assuming Gaussian errors with a root-mean-square (RMS) equal to the noise levels described in Sec. \ref{sec:mock}.  For a set of best-fit parameters $p$ we define the log–likelihood as:

\begin{equation}
\ln\mathcal{L}(p) \;=\;
-\frac12\sum_{i=1}^{N}
\frac{\bigl[F_{i}-M_{i}(p)\bigr]^2}{\mathrm{RMS}^2},
\end{equation}
where $F_i$ and $M_i(p)$ denote the observed and model flux densities in the $i$-th spectral channel, respectively, and we compute the Bayesian information criterion (BIC)
\begin{equation}
{\rm BIC} \;=\; -2\,\ln\mathcal{L} + k\,\ln N,
\end{equation}
where $k$ is the number of free parameters and $N$ the number of spectral channels. 

A broad component was considered significantly detected when at least one of the following conditions was satisfied:

\begin{enumerate}[label=(\roman*)]
\item $\Delta{\rm BIC}={\rm BIC}_{\mathrm{single}}-{\rm BIC}_{\mathrm{double}} > 10$, which indicates very strong statistical preference for the two-component model over the single-Gaussian fit, according to the scale of \citet{KassRaftery95};
\item ${\sigma}_{\mathrm{single}} > 2\,\sigma_{\mathrm{dyn}}$ (see Sec.~\ref{sec:mock}).
\end{enumerate}

This dual criterion was designed to provide a conservative yet physically motivated selection. The $\Delta{\rm BIC} > 10$ condition ensures that adding a broad component yields a statistically significant improvement to the fit. The secondary constraint, based on the line width, requires that the measured ${\sigma}_{\mathrm{single}}$ exceeds twice the gas velocity dispersion (${\sigma}_{\mathrm{single}} > 2 \sigma_{\mathrm{dyn}}$). This condition ensures that the observed broadening cannot be explained by the host galaxy’s kinematics alone and effectively excludes systems where the line width could be produced by dynamically hot gas in massive star-forming galaxies rather than by black hole–driven emission.
Cases that do not satisfy either condition are classified as non-detections. The analysis is restricted to type~1 (unobscured) AGN, and we therefore do not model additional incompleteness due to nuclear obscuration of the broad-line region.

To illustrate our fitting procedure and detection criteria, we show in Fig.\ref{fig:fit_example} a representative example of a mock spectrum with an embedded broad H$\alpha$ component. The top panel compares the single- and double-Gaussian fits to the synthetic data, while the bottom panel displays the residuals for both models. In this example, the two-component fit provides a significantly better match to the data, with a $\Delta{\rm BIC} = 52$. The broad component is unambiguously recovered in this case, with the double-Gaussian fit strongly preferred over the single-component model.
\subsection{Parameter space and detectability mapping}
\label{sec:parameterspace}

To quantify the conditions under which AGN broad-line emission is detectable, we built a four-dimensional grid of physical parameters, covering:

\begin{itemize}
    \item $M_\star = 10^8$–$10^{12}\, \rm M_\odot$ 
    \item  $\mathrm{SFR} = 0.1$–$200\, \rm M_\odot\,\mathrm{yr}^{-1}$
    \item $M_{\mathrm{BH}} = 10^6$–$10^{10}\, \rm M_\odot$ 
    \item $\lambda_{\mathrm{Edd}} = 0.1$–$1$ 
\end{itemize}

We sampled the $M_{\mathrm{BH}}$–$M_\star$ space with a $50\times50$ logarithmic grid. For each grid point, we generated 100 realizations by randomly selecting values of $\lambda_{\mathrm{Edd}}$ and SFR within their allowed ranges, assuming uniform distributions in log-space. For each realization, a synthetic H$\alpha$ spectrum was created and processed through the spectral fitting and detection procedure described in Sec.\ref{sec:fitting}.

We then computed, for each point on a grid of independently sampled $M_{\mathrm{BH}}$ and $M_\star$ values, the fraction of realizations in which the broad H$\alpha$ component was successfully detected according to the criteria defined in Sec.\ref{sec:fitting}. This quantity, which we refer to as the \textit{detection probability}, ranges from 0 (never detected) to 1 (always detected), and describes how detectability varies across the $M_{\mathrm{BH}}$–$M_\star$ plane.

This process was repeated for the two different noise levels, corresponding to 1$\times$ and 4$\times$ the continuum RMS measured in a representative JADES source (ID=73488; see Sec.~\ref{sec:mock}).
 The final results are two bi-dimensional maps, each representing the average detectability of the AGN broad-line component as a function of stellar mass and black hole mass, under different observational conditions. Fig. \ref{fig:detectability_map} illustrates these bi-dimensional detectability maps, highlighting how the observability of broad-line AGN varies with black hole and stellar mass under different noise conditions. These maps provide a direct visual insight into the regions of this parameter space where AGN signatures are likely to be recovered, and where they are suppressed due to host galaxy properties or limited sensitivity. 

These detectability maps are not only informative for understanding the observational limitations of broad-line AGN detection, but also play a fundamental role in the statistical modeling of the $M_{\mathrm{BH}}$–$M_\star$ relation. Specifically, they define the effective selection function imposed by our detection criteria and observational noise levels, acting as a truncation surface in the multi-dimensional parameter space. As a result, any inference on the intrinsic $M_{\rm BH}-M_{\star}$ relation must account for the incompleteness bias within the parameter space specific regions of the $M_{\mathrm{BH}}$–$M_\star$ plane.

In the next section, we incorporate these maps into a truncated likelihood framework to derive unbiased estimates of the parameters governing the black hole–stellar mass relation.

\subsection{Bayesian inference of the \texorpdfstring{$M_{\mathrm{BH}}$-$M_\star$}{MBH-Mstar} relation}
\label{sec:bhmassfit}

To infer the intrinsic scaling relation between black hole mass and stellar mass, we adopt a Bayesian framework that explicitly incorporates the observational selection bias described in Sec.\ref{sec:parameterspace}. Specifically, the detectability maps derived from our simulations define, for each level of observational noise, a visibility boundary in the $M_{\mathrm{BH}}$–$M_\star$ plane below which broad H$\alpha$ components are unlikely to be recovered. This implies that the observed sample of AGN is effectively censored, and any inference of the underlying scaling relation must correct for this truncation in the likelihood function.  

We model the relation as a log–linear scaling with intrinsic scatter:
\begin{equation}
\log M_{\mathrm{BH}} = \alpha + \beta\,\log\left(M_\star\right),
\end{equation}
where the free parameters of the model are the normalization $\alpha$, the slope $\beta$, and the intrinsic scatter $\sigma_{\mathrm{int, \bot}}$.  
The intrinsic dispersion is defined in the direction orthogonal to the best-fitting relation rather than vertically in $\log M_{\mathrm{BH}}$. We include observational uncertainties on both coordinates, adopting a fixed error of 0.4 dex on $\log M_\star$, following \citet{Judobaliz25}, and a conservative uncertainty of 0.4 dex on $\log M_{\mathrm{BH}}$, consistent with typical errors associated with single-epoch virial mass estimates (see, e.g., \citealt{Trefoloni25}).
For each observed galaxy, the likelihood incorporates Gaussian measurement uncertainties on both coordinates, treats the intrinsic scatter as a free model parameter, and is truncated according to the detectability boundary. The detectability map is interpolated into a smooth function $\log M_{\rm BH} = f(\log M_{\star})$, which provides the minimum detectable $\log M_{\mathrm{BH}}$ at a given stellar mass. 
Since our detectability analysis yields two representative boundaries corresponding to different noise regimes, we adopt a simplified scheme to account for the heterogeneous data quality of the compiled sample. 
For sources lying below the more conservative (higher-noise) boundary, the lower-noise limit is applied. 
For sources lying above it, the truncation boundary is randomly assigned to one of the two limits with equal probability. 
While an ideal treatment would require computing an observation-specific detectability threshold for each source, such a level of detail is not feasible for the full set of current JWST observations. 
This approach therefore provides a flexible and conservative approximation of the effective selection function across the sample.

The log-likelihood for a single data point is then
\begin{equation}
\begin{split}
\ln \mathcal{L} = \sum_{i=1}^{N} \Bigg\{ 
& -\tfrac{1}{2}\ln\!\left(2\pi\,\sigma_{\perp,i}^2\right) 
  - \tfrac{1}{2}\frac{d_{\perp,i}^2}{\sigma_{\perp,i}^2} \\
& - \ln\!\left[1-\Phi\!\left(\frac{d_{\perp,{\rm cut},i}}
    {\sigma_{\perp,i}+\sigma_{\rm soft}}\right)\right]
\Bigg\},
\end{split}
\end{equation}
where $d_{\perp,i}$ is the orthogonal distance of the $i$-th data point from the relation, 
$\sigma_{\perp,i}^2 = \sin^2\theta\,\sigma_x^2 + \cos^2\theta\,\sigma_y^2 - 2\rho\sigma_x\sigma_y\sin\theta\cos\theta + \sigma_{\rm int}^2$ is the effective variance along the orthogonal direction (with $\theta = \arctan\beta$), and $d_{\perp,{\rm cut},i}$ is the orthogonal distance to the detectability boundary. We assume independent measurement uncertainties on $\log M_\star$ and $\log M_{\mathrm{BH}}$, such that the covariance term vanishes and we set $\rho = 0$. Here, $\sigma_{\rm soft}=0.3$ acts as a softening factor that prevents an abrupt cutoff at the detectability boundary. Its value was empirically calibrated through our mock-recovery tests (see Sec. \ref{sec:recovery_test}), ensuring a faithful reproduction of the input scaling relations. The last term normalizes the Gaussian likelihood above the detectability threshold and ensures that the inference is not biased by truncation effects.  

Posterior sampling is performed using the \texttt{emcee} \citep{emcee} MCMC sampler with 32 walkers and 20,000 steps per walker. The first 5,000 steps are discarded as burn-in. We adopt a combination of hard bounds and Gaussian priors on the model parameters. Specifically, the parameters are restricted to the ranges $\alpha \in [-12,0]$, $\beta \in [-4,4]$, and $\sigma_{\mathrm{int}} \in [0.2,2.0]$ dex. The adopted prior centers are motivated by the normalization and slope reported for local SMBH–galaxy scaling relations \citep[e.g.,][]{KH13,RV15}, while the relatively large dispersions are chosen to ensure weakly informative priors that do not dominate the inference. 
Within these ranges, we impose Gaussian priors on $\alpha$, $\beta$ and $\sigma_{\mathrm{int}}$, namely\footnote{$\mathcal{N}(\mu,\sigma)$ denotes a normal distribution with mean $\mu$ and standard deviation $\sigma$.} $\alpha \sim \mathcal{N}(-4.0,2.0)$, $\sigma_{\mathrm{int}} \sim \mathcal{N}(0.5,0.7)$, and $\beta \sim \mathcal{N}(1.1,0.3)$.

We validate our implementation with two consistency checks: first, by fitting mock data drawn from a known $M_{\mathrm{BH}}$–$M_\star$ relation, where we show that our pipeline manages to recover the input parameters within the 68\% credible intervals; and second, by disabling the truncation term in the likelihood, which reproduces the standard uncorrected fit. The resulting posterior distributions and best-fit relation are presented in Sec.\ref{sec:results}.

\section{Results}\label{sec:results}
In this section, we present the results of our detectability analysis and the inferred $M_{\mathrm{BH}}$–$M_\star$ relation at $z \gtrsim 4$. We begin by examining how the observability of broad H$\alpha$ components varies across the parameter space of black hole and galaxy properties. We then use these results to test the biases on the $M_{\mathrm{BH}}$–$M_\star$ diagram introduced by NIRSpec observations. Finally, we incorporate the detectability maps into a truncated-likelihood framework when fitting the $M_{\mathrm{BH}}$–$M_\star$ relation. This approach allows us to obtain constraints on the slope, normalization, and intrinsic scatter of the scaling relation while properly accounting for selection biases. 

\subsection{Detectability mapping}
\label{sec:detectability_results}

\begin{figure*}
    \centering
    \includegraphics[width=1\textwidth]{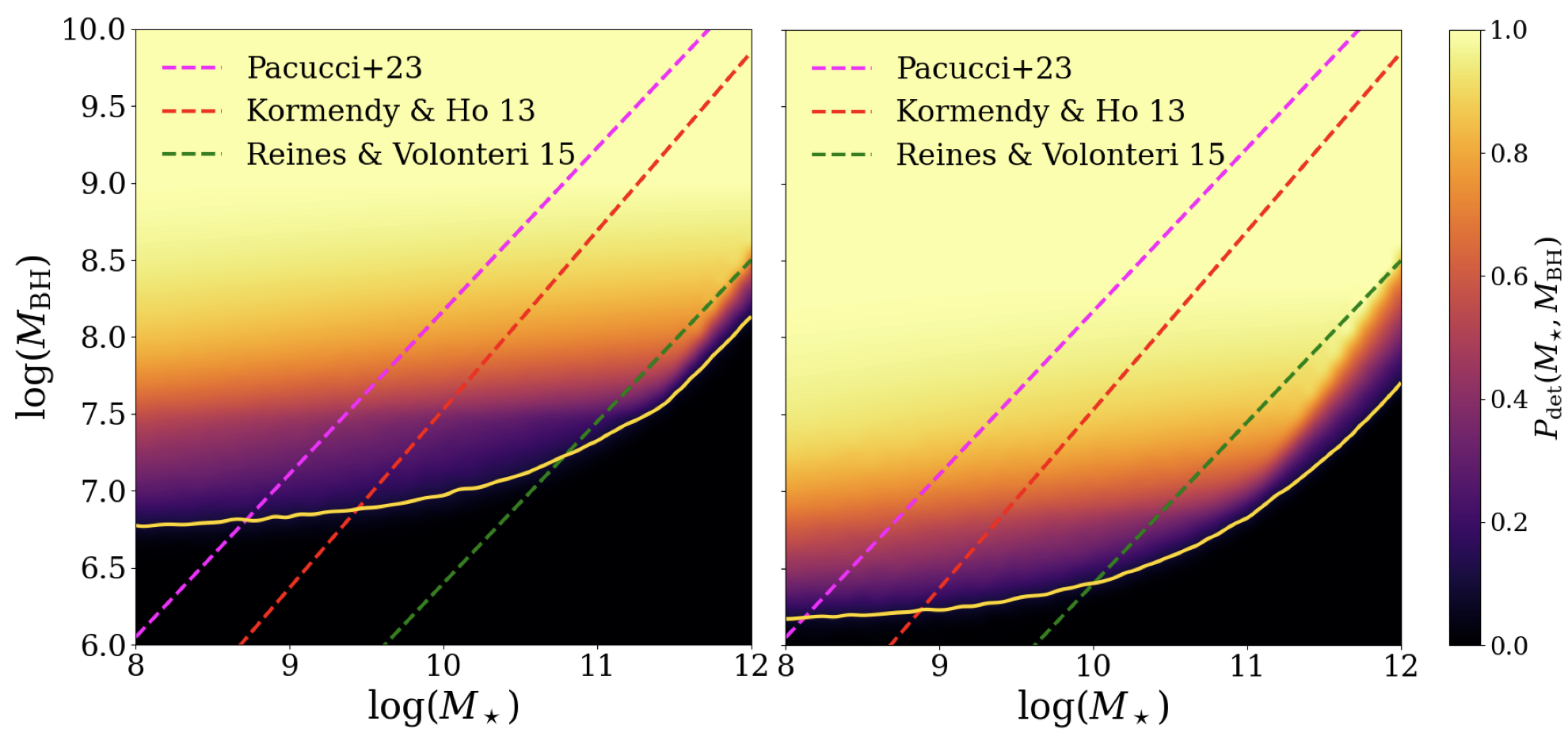}
    \caption{
  Detectability of broad H$\alpha$ emission across the $M_\star$–$M_{\mathrm{BH}}$ parameter space, for two different noise levels. Each pixel represents the average detection rate across 100 realizations with randomly sampled SFR and $\lambda_{\mathrm{Edd}}$. The color scale ranges from $0\%$ (black) to $100\%$ (yellow). The right panel assumes the lowest noise level ($1\times$ the RMS of JADES source 73488), while the left panel corresponds to the highest noise level ($4\times$ RMS). The gold continuous line indicates the detection threshold at each sensitivity, showing how reduced noise broadens the observable regime. Colored dashed lines represent $M_\star$–$M_{\mathrm{BH}}$ relations derived in previous works.
    }
    \label{fig:detectability_map}
\end{figure*}

Fig. \ref{fig:detectability_map} shows the detectability map of broad H$\alpha$ emission as a function of black hole mass and stellar mass. Each cell of the grid represents the average detection rate over 100 realizations with randomly sampled values of SFR and $\lambda_{\mathrm{Edd}}$ within the ranges defined in Sec. \ref{sec:methods}. The color scale encodes the detection probability ($P_{\rm det}$), from 0$\%$ (black) to 100$\%$ (yellow). The right panel corresponds to the deeper configuration, adopting the lower noise level of $6 \times 10^{-21} \mathrm{erg\ s^{-1}\ cm^{-2}}$ \AA\ $^{-1}$, representative of the JADES data. The left panel shows the more conservative setup, assuming the higher noise level of $2.4 \times 10^{-20}\ \mathrm{erg\ s^{-1}\ cm^{-2}}$ \AA\ $^{-1}$. The map reveals a gradual transition between detectable and non-detectable regimes, rather than a sharp threshold. In regions just above the detection boundary, broad-line components are only recovered in a subset ($\sim 30\%$) of realizations, typically those with higher $\lambda_{\rm Edd}$ or lower host galaxy SFRs. This reflects the underlying degeneracy in parameter space: the detectability of broad H$\alpha$ emission does not depend solely on BH mass or the host galaxy properties, but on their combined effect. Since each cell of the map represents the average over multiple realizations with different parameters, broad components are recovered only under favorable conditions. The uneven distribution of these favorable configurations across the $M_{\mathrm{BH}}$–$M_\star$ plane gives rise to the overall gradient in detectability and sets the overall structure of the transition between detectable and undetectable regions described below.

At fixed stellar mass, detectability increases steeply with black hole mass. In the low-mass regime (\(M_\star \lesssim 10^{10}\,M_\odot\)), even relatively small BHs ($M_{\mathrm{BH}}\sim 10^{6-7}\ M_{\odot}$) can be reliably detected, as the $H\alpha$ emission associated with the host galaxy is weaker and broad-line features stand out more clearly. In contrast, at higher stellar masses (\(M_\star \gtrsim 10^{11}\,M_\odot\)), the detection threshold shifts upward: only the most massive BHs ($M_{\mathrm{BH}}\gtrsim 10^{8}\ M_{\odot}$) remain detectable, while those with \(\log(M_{\rm BH}/M_\odot) \lesssim 7.5\) become increasingly difficult to recover due to the stronger and broader host-galaxy component. This counterintuitive behavior arises because more massive galaxies have higher gas velocity dispersions, making the broad component of small BHs comparable to the narrow component associated with the host galaxy, and thus difficult to disentangle the two components in the line fitting.

The two panels of Fig.~\ref{fig:detectability_map} also illustrate how the detectability boundary, shown as solid yellow contours corresponding to $f_{\rm det} = 0.1$, responds to different noise assumptions. The right panel corresponds to the deeper case ($1\times{\rm RMS}$), while the left panel adopts a four times higher noise level ($4\times{\rm RMS}$), representative of a shallower observational setup. As the noise decreases, the boundary shifts toward lower $M_{\mathrm{BH}}$ and $M_\star$, expanding the region where broad-line components can be reliably detected. This highlights the strong dependence of broad-line detectability on observational depth. At the high-mass end ($\log M_{\rm BH}/M_\odot \gtrsim 8.5$), the detectability remains close to $(100\%)$ across the full stellar-mass range, indicating that massive black holes are consistently recovered regardless of host properties.

\subsection{Testing the recovery of local scaling relations}
\label{sec:recovery_test}

To identify observational biases in recovering the stellar–black hole mass relation, we conducted a controlled experiment using mock galaxy populations drawn from established $z=0$ scaling relations.
Specifically, we adopted a well-established local $M_\star$–$M_{\rm BH}$ relation with its intrinsic dispersion and generated mock samples of galaxies by randomly drawing objects from the corresponding underlying distributions in the stellar mass range between $M_{\star}=10^8\ M_{\odot}$ and $M_{\star}=10^{11.5}\ M_{\odot}$, mimicking JADES observations. We performed this test twice, adopting the relations of \citet{RV15} and \citet{KH13}, respectively.
For each relation, we assigned SMBH masses by sampling from a Gaussian distribution centered on the corresponding relation, with intrinsic scatter consistent with the literature values (0.24 dex for Reines \& Volonteri, 0.3 dex for Kormendy \& Ho), noting that these values refer to the vertical scatter in $\log M_{\mathrm{BH}}$, whereas our model defines the intrinsic dispersion orthogonally to the relation. We do not treat the \citet{Greene20} relation explicitly, as it is consistent with Reines $\&$ Volonteri in normalization and probes a similar region of the black hole–stellar mass plane. We therefore expect our results to be qualitatively similar.

\begin{figure*}[t]
    \centering
    \includegraphics[width=1\textwidth]{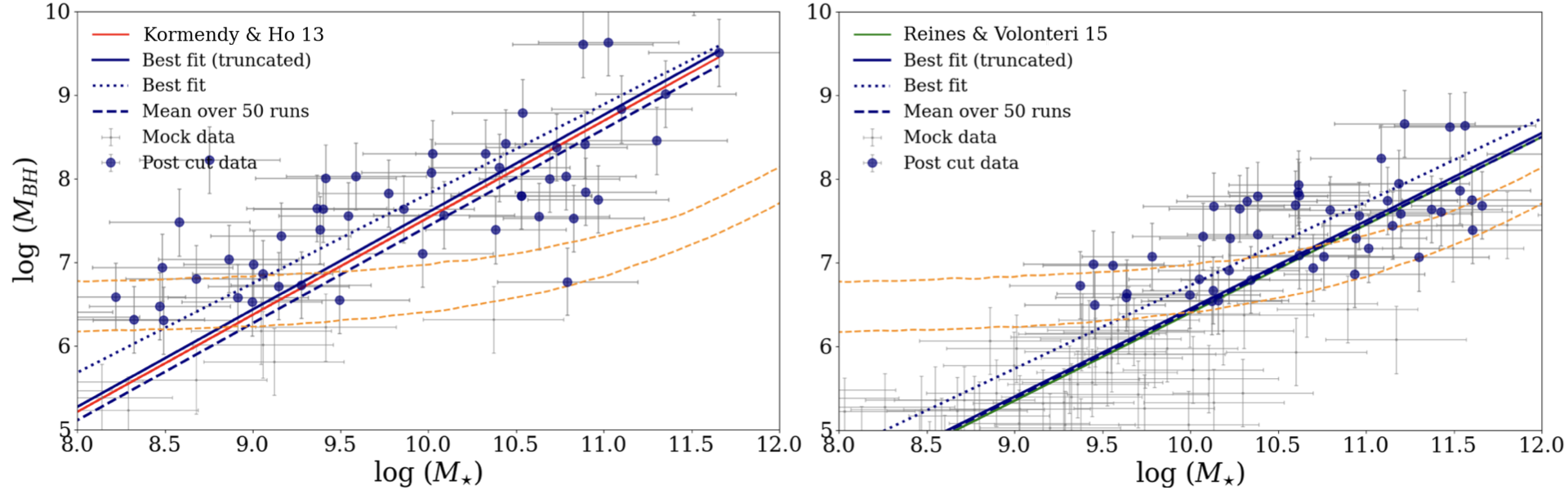}
    \caption{
Recovery test of local scaling relations under our detectability model. Each panel shows a mock sample of $\sim$50 galaxies drawn from a known $M_{\mathrm{BH}}$–$M_\star$ relation with intrinsic scatter (gray points), together with the subset of objects lying above the detectability threshold (blue points). 
The left panel adopts the \citet{KH13} relation, while the right panel assumes the lower-normalization relation of \citet{RV15}. In the latter case, the blue and green lines largely overlap and are therefore nearly indistinguishable. The orange dashed curves indicate the detectability boundaries. Solid green and red lines show the input relations, whereas the solid blue lines represent the best-fit relations obtained with our truncated-likelihood model. The blue dotted lines show the best fit obtained without accounting for truncation, and the blue dashed lines indicate the mean result over 50 realizations of our model. In both cases, our method successfully recovers the input relation within the uncertainties.}
    \label{fig:recovery_test}
\end{figure*}

We then applied the detectability cut derived from our simulations 
(Fig. \ref{fig:detectability_map}, yellow and white dashed lines), corresponding 
to the two noise levels considered. Since the JADES sample contains sources 
lying below the contour resulting from the higher noise, the most conservative 
threshold alone would not capture the observed distribution. To ensure a statistically meaningful comparison, we generated a sufficiently 
large mock sample and iteratively increased its size until 50 simulated measurements lay above the detectability threshold\footnote{This number is chosen to be comparable to the size of the observed sample ($\sim$40 objects) and slightly larger to reduce small-number fluctuations in the MC realizations.}.
We therefore combined both criteria, 
adopting the (4x) line for objects above it and the (1x) line for those below, 
so as to better reproduce the selection in the real data. Each panel in Fig.\ref{fig:recovery_test} shows the input relation (green or red), the full sample (gray points), the post-cut detectable objects (blue), and the resulting fits using a truncated likelihood method (green, as in Sec. \ref{sec:scaling_relation}).
The results clearly show that, independently of whether the underlying population follows the \citet{RV15} relation (right panel) or the \citet{KH13} relation (left panel), our methodology recovers the original trend within the statistical uncertainties after applying the detectability threshold (see Appendix$\sim$\ref{sec:app_recovery} for the full posterior distributions). In both cases, the post-cut population remains broadly consistent with the input relation, and the fitting methods reproduce the correct normalization and slope within the associated errors.

This experiment, therefore, validates our inference framework: the controlled tests demonstrate that our methodology can reliably recover the input relations within the expected uncertainties. We are thus confident that the approach can be robustly applied to the JADES sample to probe the scaling relation at high redshift.

\subsection{Inferring the \texorpdfstring{$M_{\mathrm{BH}}$-$M_\star$}{MBH–Mstar} relation}
\label{sec:scaling_relation}

\begin{figure*}[t]
    \centering
    \includegraphics[width=1.1\textwidth]{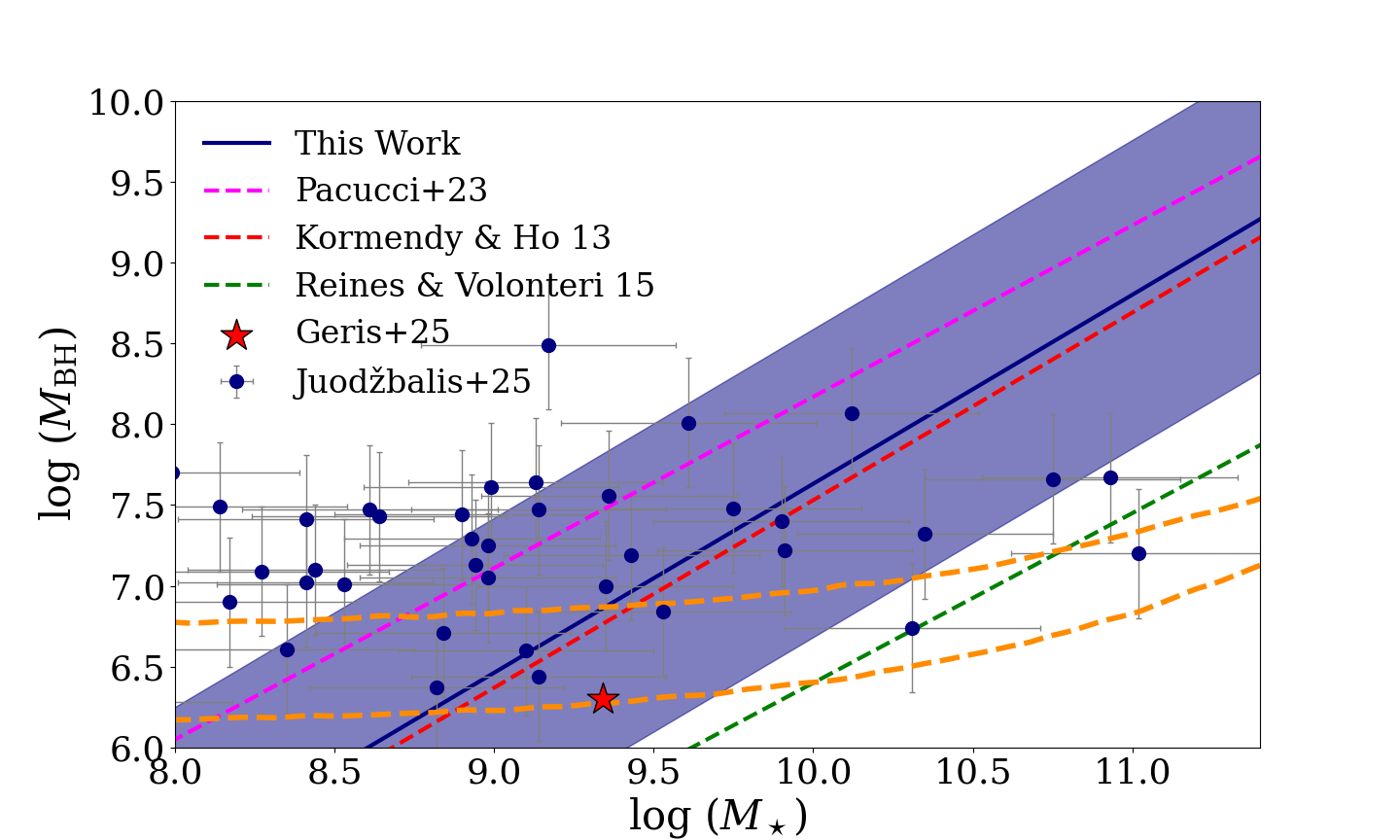}
    \caption{Black hole--host galaxy scaling relation for the \citet{Judobaliz25} sample. 
The solid blue line indicates our best-fitting $M_{\mathrm{BH}}$--$M_\star$ relation obtained with the truncated-likelihood method. 
The shaded blue region represents the intrinsic scatter ($\sigma_{\rm int}$) of the relation, 
not the uncertainty on the mean relation.  
For comparison, we show the local relations of \citet[red dashed]{KH13} and \citet[green dashed]{RV15}, 
as well as the high-redshift relation proposed by \citet[magenta dashed]{Pacucci23}. 
The two orange dashed curves mark the dynamic detectability thresholds adopted in our analysis 
(Sec.~\ref{sec:methods}).  
The red star marks the stacked measurement from \citet{Geris25}, obtained by combining $\sim$ 600 JWST/NIRSpec spectra, which probes individually undetected, low-luminosity AGN and provides a representative constraint on the low-mass end of the relation.
Overall, our inferred relation lies well above the low-normalization trend of \citet{RV15}, 
is consistent with the \citet{KH13} determination, 
and remains slightly below the \citet{Pacucci23} prediction.}
    \label{fig:scaling_relation}
\end{figure*}

Here, we present the inference of the $M_{\mathrm{BH}}$–$M_\star$ scaling relation using the broad-line AGN sample compiled by \citep{Judobaliz25} from the JADES spectroscopic survey. We restrict the analysis to sources identified within the uniform survey observations, excluding individually targeted or follow-up pointings, in order to retain a homogeneous and statistically representative population and enable a consistent modeling of the selection function and line detectability. The parent catalog already incorporates previously reported objects from other JWST programs (e.g. Early Release Observations; \citealt{Harikane23}), which are treated consistently within the same framework. This choice yields a more homogeneous and statistically representative population, avoiding cherry-picked observations and enabling a consistent modeling of the detectability of broad H$\alpha$ emission across the dataset. Fig.\ref{fig:scaling_relation} shows the distribution of sources in the $M_{\mathrm{BH}}$-$M_\star$ plane, overlaid with several reference relations \citep{KH13, RV15, Pacucci23}. Each source is associated with one of the two detection thresholds derived from the maps described in Sec.\ref{sec:methods}. The two orange dotted curves illustrate the dynamic detectability cut adopted in our analysis, combining the thresholds corresponding to different noise levels as motivated in Sec.\ref{sec:methods}.  

The best-fitting relation obtained with the truncated-likelihood model (Sec.\ref{sec:methods}), shown as the solid blue line, lies significantly above the low-normalization relation of \citet[green dashed]{RV15}, and is instead consistent within the uncertainties with the local \citet[red dashed]{KH13} relation. At the same time, it remains slightly below the high-redshift trend proposed by \citet[magenta dashed]{Pacucci23}. Overall, this result indicates that the inferred scaling relation is consistent with the local normalization reported by \citet{KH13}, which predicts relatively massive black holes, with no evidence for a further increase at high redshift. Adopting instead the lower-normalization relation of \citet{RV15} would imply a positive offset relative to the local reference.

\begin{figure*}[t]
    \centering
    \includegraphics[width=0.8\textwidth]{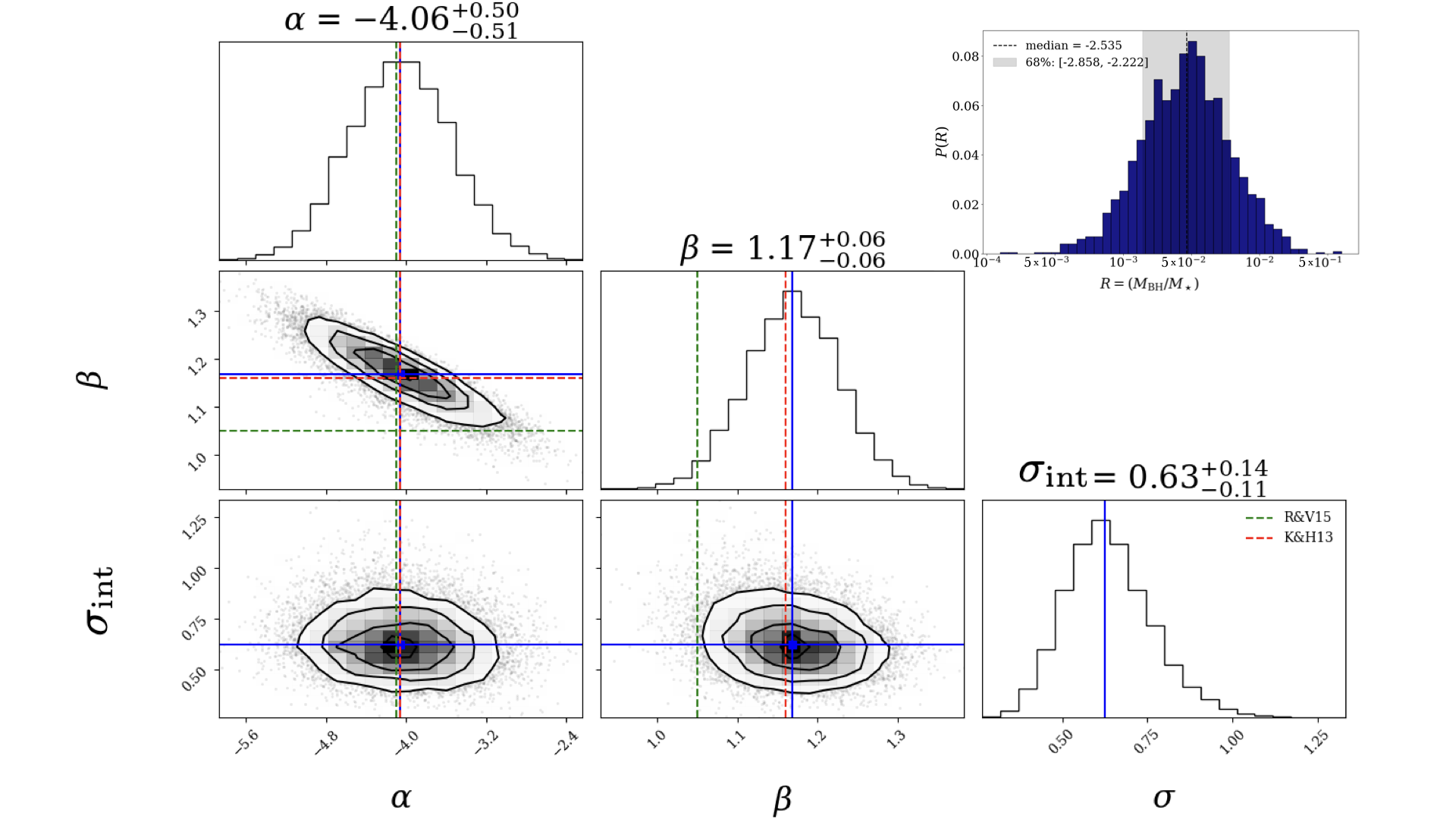}
    \caption{Posterior distributions of the parameters describing the $M_{\mathrm{BH}}$-$M_\star$ relation. 
    The panels show the marginalized one- and two-dimensional posteriors for the normalization $\alpha$, 
    slope $\beta$, and intrinsic orthogonal scatter $\sigma_{\rm int}$, as obtained from the MCMC analysis. 
    Blue lines mark the posterior medians, while red and green dashed lines indicate the values 
    corresponding to the local relations of \citet{KH13} and \citet{RV15}, respectively. 
    Contours in the 2D panels enclose the 68\% and 95\% credible regions. 
    The additional histogram in the top-right corner shows the posterior distribution of the 
    black-hole–to–stellar mass ratio, $R_{\bullet/\star} = M_{\rm BH}/M_\star$, with the shaded area marking 
    the 68\% credible interval. The comparison highlights that while the inferred slope and 
    normalization are consistent with local scaling relations, the intrinsic scatter is 
    significantly larger, indicating that the dominant evolution at $z \gtrsim 4$ lies in the 
    dispersion of the relation rather than in its mean trend.
}
    \label{fig:corner_plot}
\end{figure*}

To quantify the inferred relation, we adopt a linear parameterization of the form $\log M_{\mathrm{BH}} = \alpha + \beta \log M_\star$, with intrinsic scatter $(\sigma_{\rm int}$) defined in the direction orthogonal to the relation. The posterior distributions for these parameters, obtained from our MCMC analysis, are shown in Fig.\ref{fig:corner_plot}. We find:
\begin{equation}
    \log M_{\rm BH} = -4.06^{+0.50}_{-0.51} + 1.17^{+0.06}_{-0.06} \log M_\star,
\end{equation}
with an intrinsic orthogonal scatter of \(\sigma_{\rm int} = 0.63^{+0.14}_{-0.11}\).  

The corner plot shows the joint and marginal posterior distributions of the model parameters. The vertical dashed lines mark the values corresponding to two widely used local relations: \citet[red]{KH13} and \citet[green]{RV15}. The posterior median slope and normalization ($\alpha$) are nearly identical to those of the \citet{KH13} relation, indicating excellent agreement with this local calibration. In contrast, the best-fit slope is significantly steeper than that predicted by \citet{RV15}, which implies a flatter trend and lower normalization.
The main difference with respect to local measurements is instead observed in the intrinsic scatter ($\sigma_{\rm int}$), which is substantially larger than that inferred at low redshift. This suggests that the evolution of the $M_{\mathrm{BH}}$-$M_\star$ relation at $z \gtrsim 4$ is not driven by a systematic shift of the mean relation, but rather by an increase in its dispersion. The enhanced scatter points to a broader diversity in the growth histories of SMBHs and their host galaxies, consistent with a less settled phase of black hole galaxy co-evolution in the early Universe.

The top-right panel shows the posterior distribution of the 
black-hole–to–stellar mass ratio, \(R_{\bullet/\star} = M_{\rm BH}/M_\star\). 
The median value, \(\log R_{\bullet/\star} \simeq -2.5\), lies between the expectations from 
local relations and high-redshift predictions, with a 68\% credible interval 
spanning \([-2.8, -2.2]\). This relatively wide range reflects the large intrinsic 
scatter and illustrates the diversity of BH-to-host mass ratios at early cosmic times. The posterior distributions also show minimal degeneracy between \(\beta\) and 
\(\sigma\), demonstrating that the data constrain both the slope and the 
intrinsic dispersion. The inclusion of detectability cuts in the likelihood 
ensures that the recovered normalization is not artificially boosted by 
selection effects, representing a key improvement over previous analyses that neglected modeling detectability effects.

Overall, this analysis provides a statistically robust characterization of the 
$M_{\mathrm{BH}}$-$M_\star$ relation at $z \gtrsim 4$, consistent with local 
scaling laws in slope and normalization, but with clear evidence for increased 
intrinsic scatter compared to the nearby Universe.

\section{Discussion}\label{sec:discussion}
Motivated by the larger intrinsic dispersion we infer at $z \sim 6$, we discuss the physical mechanisms that broaden the $M_{\mathrm{BH}}$–$M_\star$ relation at early times. The variance of the relation is expected to arise from three main contributions: stochastic BH accretion and feedback, the hierarchical assembly of galaxies through mergers, and any residual diversity in the initial BH seed masses. At high redshift all three mechanisms contribute positively to the variance, naturally producing a larger intrinsic scatter even in the absence of a systematic offset from the local relation reported by \citet{KH13}.

At early epochs the intrinsic dispersion of the $M_{\rm BH}$–$M_\star$ relation is expected to be dominated by BH growth. Cosmological simulations show that stochastic accretion and feedback provide the largest contribution to the variance of $M_{\rm BH}$ at fixed stellar mass, dominating the intrinsic scatter at $z \gtrsim 3$ \citep{ZhuSpringel25}. At these epochs, cold gas fractions are high and accretion proceeds in short, intermittent episodes characterized by large Eddington ratios ($\lambda_{\rm Edd}\gtrsim0.5$) and luminous duty cycles of only $\sim10^{6-7}$ yr \citep{Hopkins05,Sijacki15,Habouzit22}. Because gas inflow to the nuclear ($\lesssim 100$ pc) region is regulated by local cooling and dynamical instabilities rather than by the global galaxy potential \citep{Ciotti10,Gaspari13,AnglesAlcazar17}, the instantaneous BH growth rate can vary by orders of magnitude on Myr timescales even among galaxies with similar stellar masses. Consequently, BHs of similar stellar mass undergo markedly different recent growth histories, temporarily appearing over-massive or under-massive relative to the mean relation and thereby increasing the intrinsic dispersion.

In contrast to accretion-driven variability, hierarchical merging acts primarily as a statistical averaging process that progressively reduces the intrinsic dispersion of the relation. In the absence of significant gas accretion, repeated mergers simply add BH and stellar masses, so that successive assembly events tend to wash out object-to-object differences and drive the population toward the mean value of the relation. Idealized merger-only models show that the scatter decreases systematically with the cumulative number of mergers ($m$), approximately following $\sigma_{\rm mrg}(m) \simeq \sigma_{\rm ini}(m+1)^{-a/2}$ with $a \sim 0.4$–0.6 over a broad range of merger histories, largely independent of the initial distribution or mass ratios \citep{Hirschmann10}. For typical merger counts $m \lesssim 50$, this scaling implies a reduction of the intrinsic scatter by a factor of $\sim 2$, such that a present-day dispersion of $\sim 0.3$ dex naturally corresponds to $\sim 0.5$–0.6 dex at $z \sim 3$–6. At high redshift, where galaxies have experienced only a limited number of assembly events, this population-wide averaging remains inefficient, and the $M_{\rm BH}$–$M_\star$ relation therefore retains a comparatively large intrinsic scatter.

A third contribution to the intrinsic dispersion is set by the initial conditions of the BH population. If BHs are seeded over a broad mass range ($\sim 10^{2}$–$10^{5},M_\odot$; \citealt{Madau98,Ferrara14,Lupi14}), this diversity directly establishes a non-zero scatter in $M_{\rm BH}$ already at formation, independently of subsequent growth. In this sense, the seed mass distribution acts as an initial variance floor: even galaxies with identical stellar masses may begin with systematically different BH masses simply because they originate from different seeds. Numerical experiments show that this imprint can represent a non-negligible fraction of the total scatter at early times, before accretion-driven regulation and repeated assembly events have had time to dominate the evolution \citep{ZhuSpringel25,Bhowmick25}. As a result, during the first $\lesssim 1$ Gyr of cosmic time, residual seed-to-seed differences can still contribute appreciably to the observed dispersion, while at later epochs their impact becomes progressively subdominant. Measurements of the high-redshift intrinsic scatter therefore offer a direct probe of the breadth of the underlying seed mass function.

\section{Comparison with previous works}\label{sec:comparison}
Several recent studies have investigated the evolution of the $M_{\mathrm{BH}}$–$M_\star$ relation in the \textit{JWST} era. In this section, we compare our results with the most relevant works, focusing in particular on how different treatments of observational selection effects impact the inferred scaling relations. A key aspect of our approach is the explicit forward modeling of the H$\alpha$ detectability, which enables a quantitative assessment of how detection biases shape the observed distribution of broad-line AGN in the ($\log M_\star$, $\log M_{\mathrm{BH}}$) plane.

A recent analysis of JWST-identified broad-line AGN at $3<z<7$ has investigated the $M_{\mathrm{BH}}$–$M_\star$ relation within a Bayesian truncated-likelihood framework \citep{Jones+25}, building on earlier methodological developments \citep{Pacucci23}. This study is methodologically similar to ours in several respects, including the use of single-epoch virial black hole mass estimates based on the broad H$\alpha$ line and an explicit attempt to account for observational selection effects, making it a natural point of comparison. A key difference, however, lies in the modeling of the selection function. In that analysis, the truncation of the likelihood is implemented through a fixed lower threshold in black hole mass, motivated by the sensitivity of JWST to broad H$\alpha$ emission. In contrast, our analysis models the detectability of the broad H$\alpha$ component explicitly through forward modeling, resulting in a selection function that varies continuously across the ($\log M_\star$, $\log M_{\mathrm{BH}}$) parameter space. Because the detectability of broad emission lines depends on multiple physical properties, such as host galaxy mass and accretion state, a fixed lower mass threshold represents a simplified approximation that can bias the inferred relation.

\begin{figure}
    \centering
    \includegraphics[width=\columnwidth]{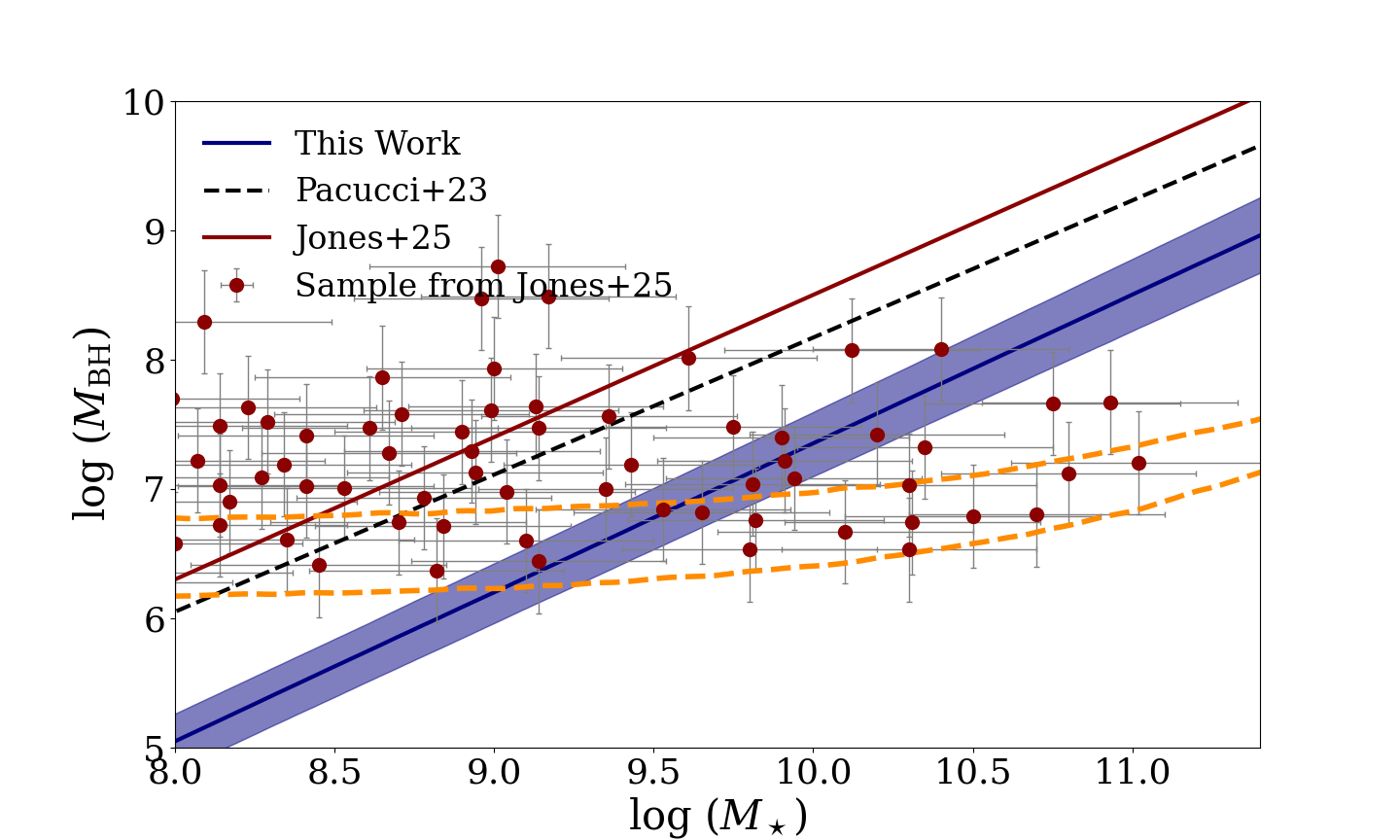}
    \caption{
    Comparison between different inferences of the $M_{\mathrm{BH}}$–$M_\star$ relation obtained from the same dataset.
    Dark red points show the sample of broad-line AGN analyzed in \citep{Jones+25}, with representative uncertainties.
    The solid dark red line indicates the best-fit relation reported in that work, while the dashed black line shows the relation inferred using an alternative truncated-likelihood approach \citep{Pacucci23}.
    The solid blue line and shaded region represent the median and 16th–84th percentile credible interval of the relation inferred in this work by applying our forward-modeled selection function to the same sample.
    The orange dashed curves indicate the effective detection boundaries adopted in our analysis.
    The shaded region reflects the uncertainty on the mean relation and does not include the intrinsic scatter.
    }
    \label{fig:jones_comparison}
\end{figure}

The two studies also adopt partially different samples. While our primary analysis is based on an independent dataset, we have carried out a direct comparison by incorporating the sources from \citep{Jones+25} into our compilation and removing duplicate objects. Applying our forward-modeled selection function to this combined sample yields an inferred $M_{\mathrm{BH}}$–$M_\star$ relation that remains consistent with our baseline results within the uncertainties ($\alpha=-4.17\pm0.5$, $\beta=1.15\pm0.1$, $\sigma_{\rm int}=0.76\pm0.1$). This comparison is shown in Figure~\ref{fig:jones_comparison}, which illustrates how different treatments of observational selection lead to different inferred normalizations of the relation, even when applied to the same underlying data. Overall, this comparison suggests that the apparent evidence for overmassive black holes and strong evolution of the $M_{\mathrm{BH}}$–$M_\star$ relation at high redshift is closely linked to the adopted selection model, underscoring the importance of a detailed and physically motivated treatment of line detectability.

A complementary perspective is provided by the population-based forward-modeling study of \citet{Li25}, which investigates the $M_{\mathrm{BH}}$-$M_\star$ relation at $4 < z < 7$ by explicitly accounting for selection effects, measurement uncertainties, and the underlying distributions of galaxy mass and AGN accretion properties. Despite the substantially different modeling strategy, their analysis similarly finds that the observed population of apparently overmassive black holes can be interpreted as the detectable upper envelope of an underlying mass relation broadly consistent with the local normalization, with a large population of lower-mass black holes remaining undetected. Compared to \citet{Li25}, our study is based on a larger and more homogeneous spectroscopic sample from JADES and directly models the detectability of broad H$\alpha$ emission at the object level. Encouragingly, the two approaches yield consistent results within the uncertainties, both favoring an intrinsic $M_{\mathrm{BH}}$-$M_\star$ relation at high redshift that is broadly consistent with that observed in the local Universe. In particular, \citet{Li25} infer a vertical intrinsic scatter of $0.97^{+0.52}_{-0.37}$~dex in $\log M_{\mathrm{BH}}$ at fixed $M_\star$ in the low-mass regime, remarkably similar to the vertical scatter implied by our orthogonal intrinsic dispersion, $\sigma_y = \sigma_{\rm int} \sqrt{1+\beta^2} \simeq 0.96$~dex. This quantitative agreement reinforces the interpretation that the observed $M_{\mathrm{BH}}$-$M_\star$ population at high redshift can be explained by a relation with substantial intrinsic scatter combined with selection effects, rather than by a systematic evolution of the mean relation.

A further complementary comparison comes from the population-based analysis of \citet{Silverman+25}, which focuses on constraining the intrinsic scatter of the $M_{\mathrm{BH}}$--$M_\star$ relation at high redshift. In that work, the joint effects of the stellar mass function, Eddington ratio distribution, AGN duty cycle, and observational selection are modeled to reproduce the observed properties of high-redshift quasars. Their results show that models with a normalization broadly consistent with the local $M_{\mathrm{BH}}$--$M_\star$ relation and a large intrinsic scatter can naturally reproduce the observed quasar population, without requiring a systematically elevated black hole mass normalization. While \citet{Silverman+25} focus on a smaller sample of luminous, massive quasars at $z>6$, our analysis is based on a larger and more heterogeneous spectroscopic sample from JADES that primarily probes lower-mass black holes and host galaxies. The two datasets therefore probe complementary regimes in BH and stellar mass. Despite differences in sample selection and modeling strategy, both analyses reach consistent conclusions: the apparently overmassive SMBHs relative to their host galaxies can be explained as the result of selection effects acting on an underlying mass relation broadly consistent with the local normalization, provided that the intrinsic scatter is large \citep[see also][]{Roberts26}.

\section{Conclusions}\label{sec:conclusions}

We have presented a statistically rigorous determination of the 
$M_{\mathrm{BH}}$--$M_\star$ relation at $z \gtrsim 4$, based on a sample of broad-line AGN from the JADES survey and a selection-aware modeling framework. By combining detectability maps with a truncated-likelihood MCMC fit, we explicitly account for the non-uniform probability of detecting broad H$\alpha$ emission. This approach allows us to mitigate selection effects that can bias black hole mass estimates at fixed host stellar mass, leading to a more reliable characterization of the underlying relation. 

Our main result is that the inferred slope and normalization of the relation are consistent with the local determination of \citet{KH13}, and significantly above the lower-normalization relation of \citet{RV15}. The best-fit relation lies below the high-redshift trend proposed by \citet{Pacucci23}. Crucially, however, we find clear evidence for a much larger intrinsic scatter than in local studies, in agreement with recent works \citep{Li25, Silverman+25}. This suggests that the dominant form of evolution in the $M_{\mathrm{BH}}$-$M_\star$ connection at early cosmic times is not a systematic shift of the mean relation, but rather an increase in dispersion. A higher scatter naturally points to a greater diversity in SMBH and galaxy growth pathways in the young Universe, where co-evolutionary processes are expected to be more stochastic and less tightly coupled than at low redshift.  

Methodologically, our analysis demonstrates the importance of explicitly 
incorporating detectability effects into scaling-relation studies. By combining mock completeness maps with a truncated-likelihood framework, we show that it is possible to recover unbiased constraints even from small, heterogeneous samples. This approach can be straightforwardly extended to future deep spectroscopic surveys, and will be essential for deriving robust constraints on SMBH–galaxy scaling laws in the high-redshift regime.  

Finally, we note that our results are based on a relatively small sample, and uncertainties remain significant. Larger samples of broad-line AGN at 
$z \gtrsim 4$, combined with improved stellar mass estimates and complementary rest-UV diagnostics, will be required to confirm the increased scatter and to map its evolution with redshift. Upcoming JWST programs, together with future facilities such as \textit{Euclid}, the \textit{Roman Space Telescope}, will provide the necessary data to refine these 
constraints.  

In summary, our findings suggest that the $M_{\mathrm{BH}}$-$M_\star$ relation was already in place by $z \sim 4-6$ at a level consistent with the local \citet{KH13} relation, but with substantially larger intrinsic scatter. 

\bibliographystyle{aa_url}
\bibliography{biblio}

\begin{appendix}

\section{Posterior distributions from the recovery tests}\label{sec:app_recovery}

\begin{figure}[t]
    \centering
    \includegraphics[width=0.5\textwidth]{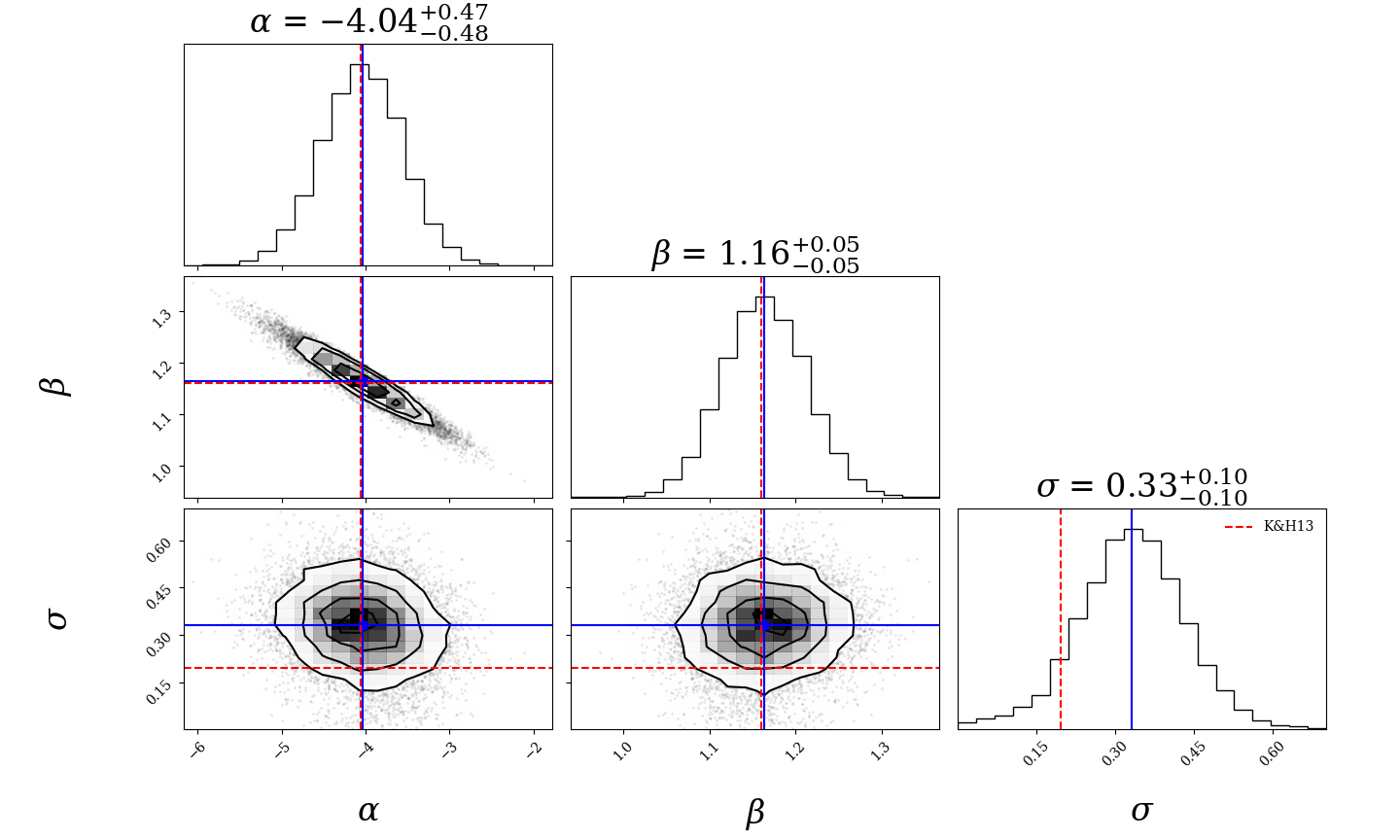}
    \caption{Posterior distributions of the parameters from the mock recovery test based on the \citet{KH13} relation.
Panels show the marginalized one- and two-dimensional posteriors for $\alpha$, $\beta$, and $\sigma$. Blue lines mark the posterior medians, while red dashed lines indicate the input values of the \citet{KH13} relation. Contours in the 2D panels enclose the 68$\%$ and 95$\%$ credible regions. The input parameters are accurately recovered, confirming the reliability of the fitting procedure.
   }
    \label{KH_Corner}
\end{figure}

\begin{figure}[t]
    \centering
    \includegraphics[width=0.5\textwidth]{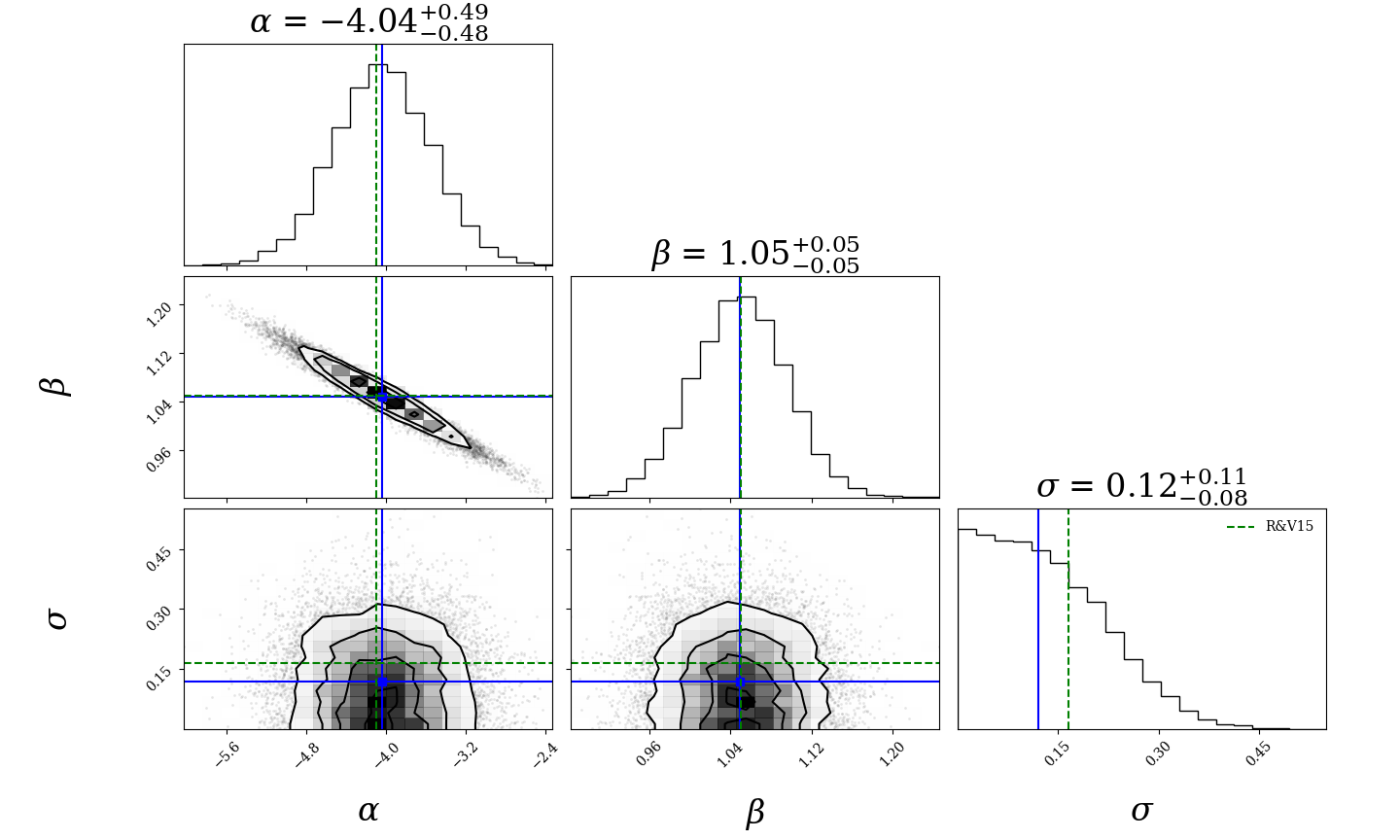}
    \caption{Posterior distributions of the parameters from the mock recovery test based on the \citet{RV15} relation.
The panels show the marginalized one- and two-dimensional posteriors for the normalization $\alpha$, slope $\beta$, and intrinsic scatter $\sigma_{\rm int}$ obtained from the MCMC analysis. Blue lines mark the posterior medians, while green dashed lines indicate the input values of the \citet{RV15} relation used to generate the mock sample. Contours in the 2D panels enclose the 68$\%$ and 95$\%$ credible regions. The recovered posterior distributions match the input relation within the statistical uncertainties.
   }
    \label{RV_Corner}
\end{figure}

In Sect.$\sim$\ref{sec:recovery_test} we presented the controlled experiments in which mock samples were generated from the \citet{RV15} and \citet{KH13} relations and subjected to the same detectability cuts adopted for the JADES data. The main text demonstrated that the truncated-likelihood fitting approach is able to recover the input relations within the quoted uncertainties. Here we complement those results by presenting the full posterior distributions of the model parameters, displayed in Fig. \ref{KH_Corner} and Fig. \ref{RV_Corner}.

For the \citet{KH13} case (input $\alpha=-4.05$, $\beta=1.16$, $\sigma_{\rm int}=0.20$), the recovered posteriors are fully consistent with the true values. The slope $\beta$ is very well reproduced, while the normalization $\alpha$ shows broader uncertainties, as expected for samples of $\sim50$ objects spanning a limited mass range. Importantly, $\sigma_{\rm int}$ is accurately recovered, with only a minor systematic overestimate of $\sim +0.1$ dex, which is insufficient to explain the substantially larger scatter measured in the data.
A similar result is obtained for the \citet{RV15} case (input $\alpha=-4.1$, $\beta=1.05$, $\sigma_{\rm int}=0.16$). Again, the slope is well constrained around the input value, the normalization is consistent but with large error bars, and the scatter $\sigma_{\rm int}$ is retrieved essentially without bias.

These tests highlight two key points: first, the methodology does not introduce systematic shifts in either slope or normalization, although the latter remains the most weakly constrained parameter. Second, and most crucially for our science case, the scatter $\sigma_{\rm int}$ is robustly recovered even after applying stringent detectability cuts. This result underpins our main finding that the intrinsic scatter of the $M_{\mathrm{BH}}$–$M_\star$ relation at $z \gtrsim 4$ is more than a factor of two larger than in the local \citet{KH13} relation. The increase in scatter therefore reflects genuine astrophysical diversity rather than an artifact of the inference framework.

\end{appendix}

\end{document}